\documentclass[letter,11pt,DIV=12,abstract=true,numbers=noenddot,titlepage=false,twocolumn=false,draft=false,preprintnumbers]{scrartcl}
\pdfoutput=1

\makeatletter

\usepackage[usenames,dvipsnames]{color}
  \definecolor{hgreen}{rgb}{0,.3,0}
  \definecolor{hred}{rgb}{.3,0,0}
  \definecolor{hblue}{rgb}{0,0,.3}
  \definecolor{LightGray}{gray}{0.95}
  \definecolor{gray}{gray}{0.6}
\usepackage[
	    colorlinks=true,
	    linkcolor=hblue,
	    citecolor=hgreen,
	    filecolor=hblue,
	    urlcolor=hred
	    ]{hyperref}
\usepackage{mathrsfs}
\usepackage[intlimits]{amsmath}
\usepackage{amssymb}
\usepackage{slashed}
\usepackage[titletoc,title]{appendix}
\usepackage[affil-it]{authblk}
\usepackage[mono=false]{libertine}
\usepackage[numbers,sort&compress]{natbib}
\usepackage{graphicx}
\usepackage{axodraw2}
\usepackage{booktabs}
\usepackage{nicefrac}
\allowdisplaybreaks

\setcapindent{1em}
\setkomafont{captionlabel}{\bfseries}
\setkomafont{caption}{\itshape}
\numberwithin{equation}{section}

\newcommand{\isum}[1]{{\textstyle\sum\limits_{#1}}}

\DeclareOldFontCommand{\bf}{\normalfont\bfseries}{\mathbf}

\begin{document}
\renewcommand\Authands{, }

\title{\boldmath 
        Generic One-Loop Matching Conditions for Rare Meson Decays
}

\date{\today}
\author[a]{Fady~Bishara%
        \thanks{\texttt{fady.bishara@desy.de}}}
\author[b]{Joachim~Brod%
        \thanks{\texttt{joachim.brod@uc.edu}}}
\author[c]{Martin~Gorbahn%
        \thanks{\texttt{Martin.Gorbahn@liverpool.ac.uk}}}
\author[c]{Ulserik~Moldanazarova%
        \thanks{\texttt{U.Moldanazarova@liverpool.ac.uk}}}
	\affil[a]{{\large Deutsches Elektronen-Synchrotron (DESY), Notkestrasse 85, D-22607 Hamburg, Germany}}
	\affil[b]{{\large Department of Physics, University of Cincinnati, Cincinnati, OH 45221, USA}}
	\affil[c]{{\large Department of Mathematical Sciences, University of Liverpool, Liverpool, L69 7ZL, UK}}

\maketitle

\begin{abstract}
Leptonic and semileptonic meson decays that proceed via
flavour-changing neutral currents provide excellent probes of physics
of the standard model and beyond. We present explicit results for
the Wilson coefficients of the weak effective Lagrangian for these
decays in any perturbative model in which these processes
proceed via one-loop contributions. We explicitly show that our
results are finite and gauge independent, and provide Mathematica code
that implements our results in an easily usable form.
\end{abstract}
\setcounter{page}{1}

\section{Introduction\label{sec:introduction}}

Recent experimental results on lepton flavour non-universality in rare $B$-meson
decays~\cite{Aaij:2021vac} and on the anomalous magnetic moment of the
muon~\cite{Abi:2021gix} have reaffirmed and strengthened the pre-existing
tensions with the corresponding standard model (SM) predictions. In the SM, both
processes are loop-induced; hence, it is reasonable to expect that physics
beyond the standard model (BSM) to also contribute at the one-loop level if
present.
The SM contribution to $B$-meson decays is well described by the weak effective
Lagrangian~\cite{Buchalla:1995vs}. The same is true for many of the SM
extensions if they involve particles with masses above the
electroweak scale. However, matching onto the effective theory is
tedious and generally has to be repeated for every new model. The
tediousness is exacerbated if one wants to, additionally, check that the
result is gauge-independent and that all UV divergences properly cancel.

In this paper, we consider generic extensions of the SM with
vectors, scalars, and fermions with the additional
assumption that the theory is perturbatively unitary and, thus,
renormalisable~\cite{LlewellynSmith:1973yud, Cornwall:1973tb,
  Cornwall:1974km}.  Once the particle content is specified, the
resulting weak effective Lagrangian can immediately be read off. The
Wilson coefficients depend on a minimal set of physical parameters and are guaranteed to be finite and gauge independent. These properties
follow from coupling constant sum rules derived from Slavnov-Taylor
identities as outlined in Ref.~\cite{Brod:2019bro}.

As an example, consider the SM contribution to the Wilson coefficient $C_9$ in
the weak effective Lagrangian.\footnote{\label{footnote:operators}The operators we will focus on in this paper are $\mathcal{O}_7^{bs}=m_b\,\bar s\sigma^{\mu\nu}P_R b\,F_{\mu\nu}$, $\mathcal{O}_{9}^\ell=(\bar s \gamma_\mu P_L b)(\bar{\ell}\gamma^\mu\ell)$, and $\mathcal{O}_{10}^\ell=(\bar s \gamma_\mu P_L b)(\bar{\ell}\gamma^\mu\gamma_5\ell)$.
Equation~\eqref{eq:lag:5flavour} then shows that $C_9^\ell = ( C_{LL}^{23\ell}+C_{LR}^{23\ell})/2$, for example. Note that we use an effective Lagrangian, as opposed to an effective Hamiltonian as in Ref.~\cite{Buchalla:1995vs}.}
Generically, the minimal field content in the loop that is required to obtain a non-zero, finite, result consists of two massive vector bosons, one charged and one neutral, two charged fermions, and one neutral fermion -- see the left panel in Table~\ref{tab:sm-example}.
Once the couplings of these states are specified, and the sum rules among them are applied, Eq.~\eqref{eq:cv-vector} directly gives the finite and gauge-independent result,
\begin{equation}
C_9 = \frac{e^2 G_F V_{ts}^*V_{tb}}{\sqrt{2}}
\left[\frac{1}{s_W^2}  F_{V}^{L,BZ}
                   - 4 F_{V}^{\gamma Z}\right]\,,
\end{equation}
where $F_V^{L,BZ}$ and $F_V^{\gamma Z}$ are loop functions that, in the SM, only depend on $m_t^2/m_W^2$, see Eq.~\eqref{eq:sm-c9-c10}.
Here, $G_F$ is the Fermi constant, $e$ is the positron charge, $s_W \equiv \sin\theta_W$ is the sine of the weak mixing angle, and $V_{ij}$ are the elements of the Cabibbo-Kobayashi-Maskawa matrix.
The procedure is exactly the same for any extension of the SM, it's that simple!

There are two important points to note here. First, the unitarity of the quark-mixing matrix is guaranteed by the sum rule in Eq.~\eqref{eq:unitarity_sumrule}.
Furthermore, in the absence of tree-level flavour-changing neutral currents (FCNCs), at least two fermion generations in the loop are required to give a non-zero contribution.
Second, and more remarkable, the same sum rule, Eq.~\eqref{eq:unitarity_sumrule}, fixes the couplings of the $Z$ boson to the internal and external fermions and, consequently, it is not necessary to specify them in the first place.
In this way, the $Z$ penguin, photon penguin, and boxes are combined into gauge-independent loop functions that generalise the penguin-box expansion of Ref.~\cite{Buchalla:1990qz}.
The penguin-box functions -- $X$, $Y$, and $Z$ of Ref.~\cite{Buchalla:1990qz} -- are directly related to our functions $F_V^{L,B^{\prime} Z}$, $F_V^{L,BZ}$ and $F_V^{\gamma Z}$ in the SM limit.
Apart from an overall normalisation, the only difference is that $F_V^{\gamma Z}$ also incorporates the light particle contribution in the matching procedure.
Our functions generalise $X$, $Y$, and $Z$ to extensions of the SM with an arbitrary number of massive vectors, scalars, and fermions while remaining gauge independent.

\begin{table}[]
	\parbox{0.49\columnwidth}{\centering
	\begin{tabular}{@{\hspace{.3em}}c@{\hspace{.4em}}|@{\hspace{.4em}}c@{\hspace{.4em}}|@{\hspace{.4em}}c@{\hspace{.4em}}}\toprule[1.5pt]
		Field &  Mass & $U(1)_Q$ Charge\\
		\midrule
		$W$ & $m_W$ & $+1$\\
		$Z$ & $m_Z$ & $0$\\
		$\nu$ & $0$ & $0$ \\
		$\{u,t\}$ & $\{0, m_t\}$ & $\nicefrac{+2}{3}$ \\
		\bottomrule[1.5pt]
	\end{tabular}}\hfill
	\parbox{0.49\columnwidth}{\centering
	\begin{tabular}{c|c}\toprule[1.5pt]
		Coupling & Value\\
		\midrule
		$\{W,\,\bar t,\, b\}$ & $\nicefrac{-1}{\sqrt{2}}\,g\,U_{tb}$\\
		$\{W^*,\,\bar{s},\, t\}$ &$\nicefrac{-1}{\sqrt{2}}\,g\,U_{ts}^*$\\
		$\{W,\,\bar\nu,\, \mu\}$ &$\nicefrac{-1}{\sqrt{2}}\,g$\\
		\bottomrule[1.5pt]
	\end{tabular}}\hfill
	\caption{The loop field content (left table) and the couplings of those fields (right table). The matrix $U_{ij}$ is the two-generation quark-mixing matrix.  Note that since we only consider two fermion generations inside the loop, the charged vector couplings need to be specified only for one generation -- see text for details.}
	\label{tab:sm-example}
\end{table}

General expressions for the photon dipole have already been presented in Refs.~\cite{Bobeth:1999ww, He:2002pva, Lavoura:2003xp, Kamenik:2009cb}, while contributions of heavy new scalars and fermions to the $b \to s \ell \ell$ transition were considered in Refs.~\cite{Gripaios:2015gra,Arnan:2016cpy, Arnan:2019uhr}.
Here, we extend the discussion to the contributions of the photon and $Z$ penguins to the semileptonic current-current operators, with a special focus on proving gauge invariance in the presence of heavy vectors, and eliminating couplings to unphysical scalars such as would-be Goldstone bosons.
Moreover, we provide easy-to-use code to obtain the Wilson coefficients in general perturbatively unitary models, it is available at
\begin{center}
	\url{https://wellput.github.io}\,.
\end{center}

The paper is organized as follows. The generic interaction Lagrangian of the extended field content is given and discussed in Sec.~\ref{sec:generic-lagrangian}. The relevant sum rules are discussed in Sec.~\ref{sec:wilcos} along with the dipole and current-current Wilson coefficients. There, we also explain the cancellation of the gauge dependent terms. In Sec.~\ref{sec:pheno-applications}, we apply our setup to three models taken from the literature to illustrate how the one-loop matching contributions can be easily obtained.
We conclude and summarize our work in Sec.~\ref{sec:summary-conclusions} and give explicit expressions for the loop functions in App.~\ref{sec:loop-functions}. The additional sum rules required for the renormalisation of the $Z$ penguin are collected in App.~\ref{sec:relevant-sum-rules}.

\section{Generic Model and Effective Lagrangian}
\label{sec:generic-lagrangian}

The goal of this work is to provide the explicit form of the effective Lagrangian relevant for leptonic, semileptonic, and radiative $B$, $B_s$, and $K$ meson decays for a generic renormalisable model.
We write the five-flavour effective Lagrangian that describes the $d_j \to d_i$ transition, obtained by integrating out the $W$ and $Z$ bosons, the top quark, as well as all heavy new particles at the electroweak scale, as
\begin{equation}\label{eq:lag:5flavour}
\begin{split}
  \delta \mathcal{L}_{\Delta F =1} &= \frac{1}{16\pi^2} 
  \sum_{\substack{\ell \in \{e,\mu,\tau\}\\  \sigma, \sigma' \in \{L,R\}}} 
  C_{\sigma\sigma'}^{ij \ell}
  \left(\bar{d}_i \gamma^{\mu} P_{\sigma} d_j\right)
  \left(\bar{\ell} \gamma_{\mu} P_{\sigma'} \ell \right)\\
 & \quad +  \frac{1}{16\pi^2} \sum_{\sigma \in \{L,R\}}
    D_{\sigma}^{ij} \bar{d}_i \sigma^{\mu\nu} P_\sigma d_j
     F_{\mu\nu} + \text{h.c.}\,.
\end{split}
\end{equation}
The operators in the first sum have the form of a product of a leptonic current and a FCNC. The second sum contains the photon dipole operators. Here, $d_i = d,s,b$ denote the down-type quark fields and $\ell$ the lepton fields. 
$P_{L} \equiv (1-\gamma_5)/2$ and $P_{R} \equiv (1+\gamma_5)/2$ are the chirality projection operators, and $\sigma$ and $\sigma'$ denote the chiralities of the incoming quarks and leptons.
We neglect all operators with mass dimension larger than six.
The explicit results for the Wilson coefficients are given below in Eqs.~\eqref{eq:dipole:formula} and \eqref{eq:cv-vector}~-~\eqref{eq:cv-scalar}.

In the following, we will determine the explicit form of the Wilson coefficients $C_{\sigma\sigma'}^{ij \ell}$ and $D_{\sigma}^{ij}$ for a generic interaction Lagrangian of fermions ($\psi$), physical scalars ($h$), and vector bosons ($V_\mu$) of the form (cf. Ref.~\cite{Brod:2019bro})
\begin{equation}
\label{eq:generic_lagrangian}
\begin{split}
\mathcal{L}_{\text{int}} =& \isum{f_1 f_2 s_1 \sigma}
y_{s_1 \bar f_1 f_2}^{\sigma} h_{s_1}
\bar\psi_{f_1} P_\sigma \psi_{f_2}
+ \isum{f_1 f_2 v_1 \sigma}
g_{v_1 \bar f_1  f_2}^{\sigma} V_{v_1,\mu}
\bar\psi_{f_1} \gamma^\mu P_\sigma \psi_{f_2} \\[2pt]
& + \tfrac{i}{6} \isum{v_1 v_2 v_3}
g_{v_1v_2v_3} \Big( V_{v_1,\mu} V_{v_2,\nu}
\, \partial^{[\mu} V_{v_3}^{\nu]} + V_{v_3,\mu} V_{v_1,\nu}
\, \partial_{\vphantom{v_2}}^{[\mu} V_{v_2}^{\nu]} + V_{v_2,\mu}
V_{v_3,\nu} \, \partial_{\vphantom{v_1}}^{[\mu} V_{v_1}^{\nu]}
\Big) \\[2pt]
& + \tfrac12 \isum{v_1 v_2 s_1}
g_{v_1v_2s_1} \, V_{v_1,\mu} V_{v_2}^{\mu} h_{s_1}
- \tfrac{i}{2} \isum{v_1 s_1 s_2}
g_{v_1s_1s_2} \, V_{v_1}^{\mu} \Big(
h_{s_1} \, \partial_\mu h_{s_2} -
\big( \partial_\mu h_{s_1} \big) \, h_{s_2} \Big) \\
&+\tfrac{1}{6}\isum{s_1 s_2 s_3}g_{s_1s_2s_3} \,h_{s_1}\,h_{s_2}\,h_{s_3}
+\tfrac{1}{24} \isum{s_1 s_2 s_3 s_4}g_{s_1s_2s_3s_4} \,h_{s_1}\,h_{s_2}\,h_{s_3}\,h_{s_4}\,,
\end{split}
\end{equation}
where $\sigma \in \{L,R\}$.
The indices $f_i$, $s_i$, and $v_i$ denote the different physical fermion, scalar, and vector fields, respectively, and run over all particles in a given multiplet of the gauge group $U(1)_\text{EM} \times SU(3)_\text{color}$.
Spinor indices are suppressed in our notation.
The non-interacting part of the Lagrangian is given by the standard kinetic terms, an $R_{\xi}$ gauge fixing term
\begin{equation}\label{eq:gaugefix_massive_vector}
\mathcal{L}_\text{fix} = - \sum_v (2 \xi_{v})^{-1} F_{\bar v} F_v\,,
\quad \quad F_v = \partial_\mu V_v^\mu - \sigma_v \xi_v M_v \phi_v ,
\end{equation}
for each massive vector, and a 't~Hooft-Feynman gauge-fixing term for
the photon field. Here, $\phi_v$ and $\xi_v$ denote the Goldstone
boson and the gauge fixing parameter associated with the vector field
$V_\mu$, while the coefficient $\sigma_v$ can have the values $\pm i$
for complex fields and $\pm 1$ for real fields. The kinetic term,
furthermore, determines the trilinear interactions with the photon
field through the covariant derivatives $D_{\mu} f = (\partial_{\mu} -
i e Q_f A_{\mu}) f$ that act on a field $f$ of charge $Q_f$.  With
this choice we have $g_{\gamma \bar{f} f}^{\sigma} = e Q_f$, $g_{v \bar{v} \gamma} = e Q_v$ and $g_{\gamma s \bar{s}} = e Q_s$, where $Q_v$ and $Q_s$ denote the charges of the vector $V_{v,\mu}$ and the scalar $h_s$, respectively, and the bar denotes the coupling with a charge conjugated fields.\footnote{The QED interaction follows from the
	kinetic terms~\cite{becchi2015slavnov}:
	\begin{equation*}
	\mathcal{L}_{\mathrm{kin}} \supset \bar{f} i \slashed{D}_{\mu} f
	- \tfrac{1}{2} \left| D_{\mu} v_{\nu} - D_{\nu} v_{\mu} \right|^2
	- \tfrac{1}{4} \left|
	F_{\mu\nu} +  i e Q_v \left(\bar{v}_{\mu} v_{\nu} - v_{\mu} \bar{v}_{\nu} \right)
	\right|^2 + (D_{\mu} h_s)^{\dagger}(D^{\mu} h_s)\,.
	\end{equation*}
}

Without additional constraints, the Lagrangian of
Eq.~(\ref{eq:generic_lagrangian}) does not describe a renormalisable
quantum field theory and cannot be used to derive predictions for
physical processes that are finite and gauge independent.
The necessary constraints arise from using the Slavnov Taylor Identities (STIs)
derived in Ref.~\cite{Brod:2019bro} from the vanishing
Becchi-Rouet-Stora-Tyutin (BRST) \cite{Becchi:1975nq,Tyutin:1975qk}
transformation of suitable vertex functions.  These STIs are
sufficient to constrain the relevant couplings for $\Delta F = 1$
flavour changing transitions that are generated at one-loop order.
In addition, the STIs determine the unphysical Goldstone couplings in
terms of the physical couplings.  For instance, the Feynman rule of
the photon interactions can be read of from the generic Lagrangian by
replacing appropriate scalar fields $s$ by $\phi$ and noting that the
STIs derived in Ref.~\cite{Brod:2019bro} imply $g_{v \bar{v} \gamma} =
g_{ \gamma \phi \bar{\phi}}$. This allows us to express all
contributions of Goldstone bosons in terms of physical
couplings. Hence, all following results include all relevant
contributions from Goldstone bosons even if only physical coupling
constants appear.

\section{Results for the Wilson coefficients}
\label{sec:wilcos}

\begin{figure}[tbp]
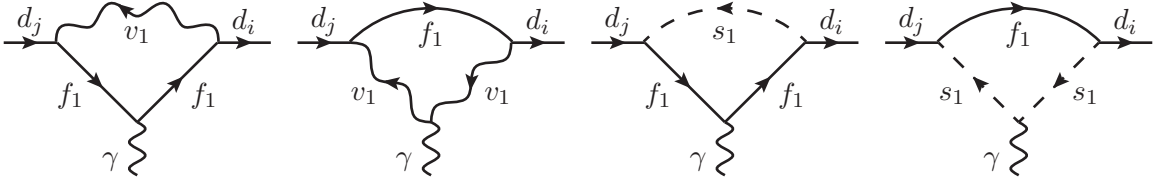

  \centering
\begin{center}
  \begin{axopicture}(430,70)
    \SetWidth{1.0}
    \SetColor{Black}
    \Text(10,58)[c]{$d_j$}
    \Text(90,58)[c]{$d_i$}
    \Text(40,05)[c]{$\gamma$}
    \Text(25,30)[c]{$f_1$}
    \Text(75,30)[c]{$f_1$}
    \Text(50,53)[c]{$v_1$}
    \Line[arrow](0,50)(20,50)
    \Line[arrow](20,50)(50,20)
    \Line[arrow](50,20)(80,50)
    \Line[arrow](80,50)(100,50)
    \Photon(50,20)(50,00){3}{2}
    \PhotonArc[clock](50,20)(43,135,45){3}{3.5}
    \Line[arrow](47,61.2)(43,64.3)
    \SetOffset(110,0)
    \Text(10,58)[c]{$d_j$}
    \Text(90,58)[c]{$d_i$}
    \Text(40,05)[c]{$\gamma$}
    \Text(25,30)[c]{$v_1$}
    \Text(75,30)[c]{$v_1$}
    \Text(50,53)[c]{$f_1$}
    \Line[arrow](0,50)(20,50)
    \Arc[clock,arrow](50,20)(43,135,45)
    \Line[arrow](80,50)(100,50)
    \Line[arrow](65.1,36)(64.9,34)
    \Photon(80,50)(50,20){3}{2}
    \Line[arrow](37,34.7)(33,35.3)
    \Photon(20,50)(50,20){3}{2}
    \Photon(50,20)(50,00){3}{2}
    \SetOffset(220,0)
    \Text(10,58)[c]{$d_j$}
    \Text(90,58)[c]{$d_i$}
    \Text(40,05)[c]{$\gamma$}
    \Text(25,30)[c]{$f_1$}
    \Text(75,30)[c]{$f_1$}
    \Text(50,53)[c]{$s_1$}
    \Line[arrow](0,50)(20,50)
    \Line[arrow](20,50)(50,20)
    \Line[arrow](50,20)(80,50)
    \Line[arrow](80,50)(100,50)
    \Photon(50,20)(50,00){3}{2}
    \Arc[dash,dashsize=7,arrow](50,20)(43,45,135)
    \SetOffset(330,0)
    \Text(10,58)[c]{$d_j$}
    \Text(90,58)[c]{$d_i$}
    \Text(40,05)[c]{$\gamma$}
    \Text(25,30)[c]{$s_1$}
    \Text(75,30)[c]{$s_1$}
    \Text(50,53)[c]{$f_1$}
    \Line[arrow](0,50)(20,50)
    \Arc[clock,arrow](50,20)(43,135,45)
    \Line[arrow](80,50)(100,50)
    \Line[dash,dashsize=7,arrow](80,50)(50,20)
    \Line[dash,dashsize=7,arrow](50,20)(20,50)
    \Photon(50,20)(50,00){3}{2}
  \end{axopicture}
\end{center}
\caption{Diagrams for the off-shell $d_j \to d_i \gamma$ Green's function.
  Physical scalars are denoted by dashed lines, while the contribution of massive vector bosons and their related Goldstone bosons are denoted by a wavy line.}
  \label{fig:photon-penguin}
\end{figure}

The Wilson coefficients of the effective Lagrangian are functions of the couplings of the generic Lagrangian and the associated masses.
They are determined by calculating suitable Green's functions:
The photon penguin diagrams (Fig.~\ref{fig:photon-penguin}) contribute in part to the dipole coefficients $D_{\sigma}^{ij}$, and in part to the current-current coefficients $C_{\sigma\sigma'}^{ij \ell}$ via the equations of motion of the photon field. 
The $Z$-penguin and box diagrams (Fig.~\ref{fig:penguin-box}) contribute to the current-current coefficients $C_{\sigma\sigma'}^{ij \ell}$.
In the remainder of this section, we spell out the details of this calculation, with a focus on obtaining a finite and gauge-independent result.

We incorporate the constraints from the STIs by repeatedly applying the sum rules on the one-loop amplitudes.
For the evaluation of the off-shell photon penguin $d_j \to d_i \gamma$ Green's function (see Fig.~\ref{fig:photon-penguin}) we only need the ``unitarity sum rule''~\cite{Brod:2019bro}
\begin{equation}
  \label{eq:unitarity_sumrule}
\sum_{v_3}\, g_{v_3 \bar d_i d_j}^{\sigma} g_{v_1 \bar{v}_2 \bar v_3} =
\sum_{f_1} \, g_{v_1 \bar d_i f_1}^{\sigma} g_{\bar{v}_2 \bar f_1 d_j}^{\sigma}
- \sum_{f_1} \, g_{\bar{v}_2 \bar d_i f_1}^{\sigma} g_{v_1 \bar f_1 d_j}^{\sigma} \,,
\end{equation}
where the summation on the right hand side of the equation is over all possible fermions $f_1$ that satisfy the charge conservation conditions.
Setting $Q_{d_i} = Q_{d_j} \equiv Q_d$ implies $Q_{v_3} = 0$ and $Q_{v} \equiv Q_{v_1} = - Q_{\bar{v}_2}$. Additionally, this implies that the charges of the fermions $f_1$ can either be $Q_{f_1} = Q_{d} - Q_{v}$ or $Q_{f_1} = Q_{d} + Q_{v}$ which respectively contribute to the first or second sum on the right hand side.
Since we only consider interactions where $g_{v_3 \bar d_i d_j}=0$ for any neutral vector $v_3$, we find the following generalisation of the Glashow-Iliopoulos-Maiani (GIM) relation
\begin{equation}
\label{eq:generalised-gim}
g^\sigma_{\bar{v}_2\bar d_i f_0} g^\sigma_{v_1\bar f_0 d_j} =
-\sum_{f_1 \neq f_0} g^\sigma_{\bar{v}_2\bar d_i f_1} g^\sigma_{v_1\bar f_1 d_j}
+ \sum_{f_1} \, g_{v_1 \bar d_i f_1}^{\sigma} g_{\bar{v}_2 \bar f_1 d_j}^{\sigma}\,.
\end{equation}
This relation can be used to eliminate the couplings of any one member of the set of fermions of charge $Q_{f_1}$ that generate flavour changing neutral currents through charged vector interactions. For definiteness, we always choose to eliminate the lightest of such fermions.
This will simultaneously determine the Wilson coefficients of the dipole operators and the photon-penguin contribution to the $\Delta F=1$ current-current operators.
The Wilson coefficients of the dipole operators are independent of the gauge fixing parameters, while the photon-penguin contribution is not.\footnote{We remark that we can project the off-shell photon Green's function onto the off-shell basis, including physical, equation-of-motion-vanishing, and BRST-exact operators, only after applying the sum rule~\eqref{eq:generalised-gim}.}

The Wilson coefficients then depend on the mass of the lightest fermion that can contribute in the loop, denoted above by the index $f_0$.
This particle could be either a light\footnote{Here, the notion of light and heavy is defined via the characteristic scale of the matching calculation, which is determined by the masses of the heavy degrees of freedom in the UV theory. In this work we assume that this is the electroweak scale, even though the formalism could be easily applied to a matching to a different effective theory.} standard-model fermion, such as an up quark, or a heavy fermion, such as a chargino.
A fermion mass of $f_0$ that is considerably smaller than the matching scale requires an appropriate effective theory counterpart that will account for the infrared logarithm generated in the limit $m_{f_0} \to 0$, see App.~\ref{sec:loop-functions-vector}.

For the renormlisation of the $Z$ penguin, two more sum rules are required~\cite{Brod:2019bro}; we list them in App.~\ref{sec:relevant-sum-rules}.

\subsection{Dipole Operator Coefficients}
\label{sec:dipole-coefficient}

Here and in the following, we write the Wilson coefficient of the five-flavour effective Lagrangian as a product of the coupling constants and gauge-independent loop functions that depend on various mass ratios defined by $x^a_b \equiv m^2_a/m^2_b$.
The matching coefficients of the dipole operators are immediately gauge independent.
We find
\begin{equation}\label{eq:dipole:formula}
\begin{split}D_{R}^{ij} &= 
  \sum\limits_{s_1f_1} \frac{ y^R_{\bar s_1\bar d_if_1}}{M_{s_1}^2}
  \left(
    m_{f_1} y^R_{s_1\bar f_1d_j} F^d_S(x^{f_1}_{s_1})
   +m_{d_j} y^L_{s_1\bar f_1d_j} F^d_{S^\prime}(x^{f_1}_{s_1})\right) \\
  & \quad + \sum\limits_{v_1f_1} \frac{g^L_{\bar v_1\bar d_if_1}}{M_{v_1}^2}
  \left(
    m_{f_1} g^R_{v_1\bar f_1d_j} F^d_{V'}(x^{f_1}_{v_1})
    +m_{d_j} g^L_{v_1\bar f_1d_j} F^d_V(x^{f_0}_{v_1}, x^{f_1}_{v_1})
  \right) \,,
\end{split}
\end{equation}
where here and in all analogous equations below the sums run over all combinations of indices that are allowed by charge and colour conservation. The explicit form of the loop functions is given in App.~\ref{sec:loop-functions-dipole}.
The first line represents the contribution of internal fermions and scalars.
The appearance of two left-handed Yukawa couplings in the first term requires an odd number of mass insertions in the fermion line, hence the loop function $F_S^d$ is multiplied with the internal fermion mass $m_{f_1}$. 
The mass factor in the second term is supplied by the Dirac equation acting on the external $d_j$ spinor (we neglect the lighter $m_{d_i}$ mass).
The second line represents the effects of internal charged massive vector bosons and fermions.
Now, the first term proportional to two vector couplings of opposite chirality requires an odd number of mass insertions, resulting in the factor $m_{f_1}$.
The second term, proportional to the function $F_V^d$, involves two vector couplings of the same chirality and receives a factor $m_{d_j}$ from the Dirac equation.
Moreover, we have used Eq.~\eqref{eq:generalised-gim}, generating the explicit dependence on the mass of the fermion $f_0$.
If there are fermions of charge $Q_{f'_1} = Q_{d} - Q_{v}$ we have to add their contribution through the sum
\begin{equation}
\label{eq:fprime}
D_{R}^{ij} \to D_{R}^{ij} +\sum\limits_{\bar v_1f'_1}
\frac{m_{d_j}}{M_{v_1}^2} g^L_{v_1\bar d_if'_1} g^L_{\bar v_1\bar f'_1d_j} F^d_{\bar{V}}(x^{f_0}_{v_1}, x^{f'_1}_{\bar v_1})
\end{equation}
if the generalised GIM mechanism of Eq.~\eqref{eq:generalised-gim} has already been applied to the sum of fermions of charge $Q_{f_1} = Q_{d} + Q_{v}$.
The modified loop function $F^d_{\bar{V}}$ is obtained from $F^d_{V}$ by the simple replacement $Q_{v_1} \to -Q_{v_1}$.
Finally, let us note that we could further simplify the function $F^d_{V'}$ using the sum rule Eq.~\eqref{eq:gauge_mass_sumrule} if tree-level neutral current and scalar interactions are absent.
In this limit we have
\begin{equation}
\label{eq:generalised-left-right-gim}
m_{f_0} g^R_{v_1\bar f_0 d_j}g^L_{\bar{v}_2\bar d_i f_0} =
-\sum_{f_1 \neq f_0}
m_{f_1} g^R_{v_1\bar f_1 d_j}g^L_{\bar{v}_2\bar d_i f_1} \,,
\end{equation}
when we set $m_{d_j} = m_{d_i} = 0$.
Our results agree with Ref.~\cite{Lavoura:2003xp} if we apply our generalised GIM mechanism to their results.
Here we note that it is only possible to project the off-shell Green's function after using the GIM mechanism.
The coefficients $D_{L}^{ij}$ can be recovered from $D_{R}^{ij}$ by simply interchanging the chirality of all coupling constants, i.e.\ by replacing $y^L_{\cdots} \leftrightarrow y^R_{\cdots}$ and $g^L_{\cdots} \leftrightarrow g^R_{\cdots}$ in Eq.~(\ref{eq:dipole:formula}).

\subsection{Neutral-Current Wilson Coefficient}
\label{sec:neutral-current-coefficient}

\begin{figure}[tbp]
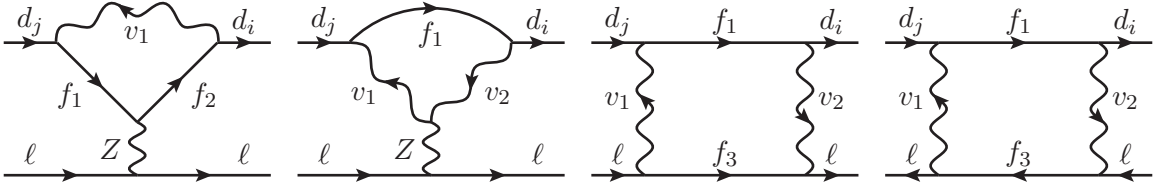

  \centering
\begin{center}
  \begin{axopicture}(430,70)
    \SetWidth{1.0}
    \SetColor{Black}
    \Text(10,58)[c]{$d_j$}
    \Text(90,58)[c]{$d_i$}
    \Text(40,10)[c]{$Z$}
    \Text(25,30)[c]{$f_1$}
    \Text(75,30)[c]{$f_2$}
    \Text(50,53)[c]{$v_1$}
    \Line[arrow](0,50)(20,50)
    \Line[arrow](20,50)(50,20)
    \Line[arrow](50,20)(80,50)
    \Line[arrow](80,50)(100,50)
    \Photon(50,20)(50,00){3}{2}
    \PhotonArc[clock](50,20)(43,135,45){3}{3.5}
    \Line[arrow](47,61.2)(43,64.3)
    \Text(10,8)[c]{$\ell$}
    \Text(90,8)[c]{$\ell$}
    \Line[arrow](0,0)(50,0)
    \Line[arrow](50,0)(100,0)
    \SetOffset(110,0)
    \Text(10,58)[c]{$d_j$}
    \Text(90,58)[c]{$d_i$}
    \Text(40,10)[c]{$Z$}
    \Text(25,30)[c]{$v_1$}
    \Text(75,30)[c]{$v_2$}
    \Text(50,53)[c]{$f_1$}
    \Line[arrow](0,50)(20,50)
    \Arc[clock,arrow](50,20)(43,135,45)
    \Line[arrow](80,50)(100,50)
    \Line[arrow](65.1,36)(64.9,34)
    \Photon(80,50)(50,20){3}{2}
    \Line[arrow](37,34.7)(33,35.3)
    \Photon(20,50)(50,20){3}{2}
    \Photon(50,20)(50,00){3}{2}
    \Text(10,8)[c]{$\ell$}
    \Text(90,8)[c]{$\ell$}
    \Line[arrow](0,0)(50,0)
    \Line[arrow](50,0)(100,0)
    \SetOffset(220,0)
    \Text(10,58)[c]{$d_j$}
    \Text(50,58)[c]{$f_1$}
    \Text(90,58)[c]{$d_i$}
    \Line[arrow](0,50)(20,50)
    \Line[arrow](20,50)(80,50)
    \Line[arrow](80,50)(100,50)
    \Text(10,8)[c]{$\ell$}
    \Text(50,8)[c]{$f_3$}
    \Text(90,8)[c]{$\ell$}
    \Line[arrow](0,0)(20,0)
    \Line[arrow](20,0)(80,0)
    \Line[arrow](80,0)(100,0)

    \Text(10,28)[c]{$v_1$}
    \Line[arrow](21,27.5)(19,29.5)
    \Photon(20,0)(20,50){3}{3.5}

    \Text(90,28)[c]{$v_2$}
    \Line[arrow](79,22.5)(81,20.5)
    \Photon(80,50)(80,0){3}{3.5}

    \SetOffset(330,0)
    \Text(10,58)[c]{$d_j$}
    \Text(50,58)[c]{$f_1$}
    \Text(90,58)[c]{$d_i$}
    \Line[arrow](0,50)(20,50)
    \Line[arrow](20,50)(80,50)
    \Line[arrow](80,50)(100,50)
    \Text(10,8)[c]{$\ell$}
    \Text(50,8)[c]{$f_3$}
    \Text(90,8)[c]{$\ell$}
    \Line[arrow](20,0)(0,0)
    \Line[arrow](80,0)(20,0)
    \Line[arrow](100,0)(80,0)

    \Text(10,28)[c]{$v_1$}
    \Line[arrow](21,27.5)(19,29.5)
    \Photon(20,0)(20,50){3}{3.5}

    \Text(90,28)[c]{$v_2$}
    \Line[arrow](79,22.5)(81,20.5)
    \Photon(80,50)(80,0){3}{3.5}
  \end{axopicture}
\end{center}
\caption{Diagrams that directly match onto the $\Delta F=1$ current-current operators.
  Here only contributions of internal massive vector bosons and fermions is shown.
  In addition, there is a finite contribution from the off-diagonal fermion self-energy diagram.
  The contribution of massive vector bosons and their related Goldstone bosons are denoted by a wavy line.}
  \label{fig:penguin-box}
\end{figure}

Both the photon penguin diagrams of Fig.~\ref{fig:photon-penguin} and the $Z$ penguin and box diagrams of Fig.~\ref{fig:penguin-box} contribute to the matching conditions for the current-current Wilson coefficients.
The analytic expression of each of the three diagram classes depends on the gauge fixing parameters of the massive vector bosons in the loop.
A renormalised result for the $Z$ penguin was derived in Ref.~\cite{Brod:2019bro} in 't~Hooft-Feynman gauge using sum-rules derived from Slavnov-Taylor identities.
Here we will show how to apply these same sum rules to combine the amplitudes of all three diagram classes into a finite and gauge-parameter independent result for the Wilson coefficients. To this end, we split our final expression into three parts,
\begin{equation}
  \label{eq:CVLL-formula}
  \tilde{C}_{L \sigma}^{i j \ell}
  = v_{L \sigma}^{i j \ell} + m_{L \sigma}^{i j \ell} + s_{L \sigma}^{i j \ell} \,,
\end{equation}
as a sum of diagrams that in the loop contain only massive vectors and fermions, denoted by $v_{L \sigma}^{i j \ell}$, massive vectors, massive scalars and fermions, denoted by $m_{L \sigma}^{i j \ell}$, and massive scalars and fermions, denoted by $s_{L \sigma}^{i j \ell}$.
The index $L$ denotes the left chirality of the external quarks, while $\sigma=L,R$ stands for the chirality of the external field $\ell$.
Again, the expressions for $\tilde{C}_{R \sigma}^{i j \ell}$ can be recovered from $\tilde{C}_{L \sigma}^{i j \ell}$ by simply swapping the chirality of all coupling constants, i.e.\ by replacing $y^L_{\cdots} \leftrightarrow y^R_{\cdots}$, $g^L_{\cdots} \leftrightarrow g^R_{\cdots}$ and $\sigma \leftrightarrow \bar{\sigma}$, where $\bar{L}=R$ and vice versa.

The contribution of massive vectors and fermions,
\begin{equation}
  \label{eq:cv-vector}
  \begin{aligned}
    v_{L \sigma}^{i j \ell} =& \sum\limits_{v_1v_2 f_1}
    \frac{g^L_{\bar{v}_2\bar d_i f_1} g^L_{v_1\bar f_1 d_j}}{M_{v_1}^2}
    \Bigg[
    e^2 Q_\ell \delta_{v_1v_2} F_V^{\gamma Z}(x^{f_0}_{v_1},x^{f_1}_{v_1}) \\
 & \qquad + \sum\limits_{f_3} \Big(
      g^\sigma_{\bar v_1 \bar \ell f_3}g^\sigma_{v_2\bar f_3 \ell}
      F_V^{\sigma,B Z}(x^{f_0}_{v_1},x^{f_1}_{v_1},x^{v_1}_{v_2},x^{f_3}_{v_1})\\
      & \qquad \qquad \qquad +
      g^\sigma_{v_2 \bar \ell f_3}g^\sigma_{\bar v_1 \bar f_3 \ell}
      F_V^{\sigma,B' Z}(x^{f_0}_{v_1},x^{f_1}_{v_1},x^{v_1}_{v_2},x^{f_3}_{v_1})
    \Big)
    \Bigg ] \\
    + & \sum\limits_{Zv_1v_2 f_1 f_2}
    \frac{g^\sigma_{Z \bar \ell\ell} g^L_{v_1\bar f_1 d_j}g^L_{\bar{v}_2\bar d_i f_2}}{M_Z^2}
    \Bigg\{
    \delta_{f_1 f_2} g_{Z \bar{v}_1v_2} F_{V''}^{Z} (x^{f_0}_{v_1},x^{f_1}_{v_1},x^{v_1}_{v_2}) \\
    & \qquad \qquad \qquad
    + \delta_{v_1 v_2}
    \left[
    g^L_{Z \bar f_2 f_1}
    F_{V}^{Z} ( x^{f_1}_{v_1},x^{f_2}_{v_1} ) +
    g^R_{Z \bar f_2 f_1}
    F_{V'}^{Z} ( x^{f_1}_{v_1},x^{f_2}_{v_1} )
  \right]
  \Bigg\}\,,
  \end{aligned}
\end{equation}
contains several gauge-independent loop functions.
The functions $F_V^{\gamma Z}$ and $F_V^{\sigma,B^{(\prime)} Z}$ are the gauge-independent combinations of the photon penguin with the $Z$ penguin and the photon penguin with the box-diagrams.
While all of the above functions involve contributions from the lightest fermionic particle in the   loop through our generalised GIM mechanism, only $F_V^{\gamma Z}$ will contain an infrared logarithm in the limit $x^{f_0}_{v_1} \to 0$.
This logarithm is reproduced by a light-quark loop involving $f_0$ in the effective theory.
In the standard model this corresponds to the leading logarithm associated with the mixing of the operator $Q_2$ into $Q_9$ of Ref.~\cite{Buchalla:1990qz}.
The loop function $F_V^{\gamma Z}$ reproduces this leading logarithm if the considered model of new physics has the same light-particle content as the standard model.
It will then drop out in the difference of the standard model and the new-physics contribution and we can consider the resulting difference the leading new-physics contribution.

There are two gauge-independent combinations for the $Z$-penguin and box diagram that are distinguished by their fermion flow.
Charge conservation implies that the left box diagram in Fig.~\ref{fig:penguin-box} contributes if $Q_{f_3} = Q_{\ell} + Q_{d_j} - Q_{f_1}$, while the right box diagram contributes if $Q_{f_3} = Q_{\ell} - Q_{d_j} + Q_{f_1}$.
In the SM, $F_V^{\sigma,B Z}$ and $F_V^{\sigma,B^{\prime} Z}$ will then contribute to $b \to s \mu^+ \mu^-$ and $s \to d \bar{\nu} \nu$, respectively.

The loop functions $F_{V^{(\prime/\prime\prime)}}^{Z}$ are the $M_Z$-independent parts of the functions evaluated in Ref.~\cite{Brod:2019bro} and are only non-zero in physics beyond the standard model.
In particular, we note that all contributions with diagonal $Z$ couplings vanish since $F_{V^{(\prime)}}^{Z}(x,x) = F_{V''}^{Z}(x,y,1) = 0$.

Finally, we give the contributions involving internal scalars, vectors, and fermions,
\begin{equation}
  \label{eq:cm-mixed}
  \begin{split}
    m_{L \sigma}^{i j \ell} =& \hspace{-6pt} \sum\limits_{s_1v_1f_1f_3}
    \frac{1}{M^2_{v_1}}
    \big(g^L_{\bar{v}_1\bar d_if_1} y^L_{s_1 \bar f_1d_j}+ y^R_{\bar{s}_1\bar d_if_1} g^L_{v_1 \bar f_1d_j}\big)\\
    & \qquad \qquad \times
    \big( y^{\bar\sigma}_{\bar{s}_1\bar \ell f_3} g^\sigma_{v_1 \bar f_3\ell} + g^{\sigma}_{\bar v_1\bar \ell f_3} y^\sigma_{s_1 \bar f_3\ell} \big) F_{VS}^B(x^{f_1}_{s_1},x^{s_1}_{v_1},x^{f_3}_{s_1}) \\ 
    +&\hspace{-6pt} \sum\limits_{s_1v_1f_1Z} \frac{g^\sigma_{Z\bar \ell\ell}}{M_Z^2}
    \Big[ g^L_{\bar{v}_1\bar d_if_1}  y^L_{s_1 \bar f_1d_j} g_{Zv_1 \bar{s}_1} F^Z_{VS}(x^{f_1}_{s_1},x^{f_1}_{v_1})\\
    & \qquad \qquad \quad 
      + y^R_{\bar{s}_1\bar d_i f_1} g^L_{v_1 \bar f_1d_j} g_{Z\bar{v}_1s_1} F^Z_{VS^\prime}(x^{f_1}_{s_1},x^{f_1}_{v_1})
\Big]\,,
  \end{split}
\end{equation}
and only scalars and fermions,
\begin{equation}
  \label{eq:cv-scalar}
  \begin{aligned}
    s_{L \sigma}^{i j \ell} =&  \hspace{-6pt}\sum\limits_{s_1s_2f_1} \frac{1}{M^2_{s_1}}
    y^L_{s_1 \bar f_1d_j}y^R_{\bar{s}_2\bar d_i f_1} \\
    & \qquad \times \Bigg \{
    \delta_{s_1s_2}  e^2 Q_{\ell} F_S^\gamma(x^{f_1}_{s_1}) +
     \sum\limits_{f_3} \left( y^{\bar\sigma}_{\bar{s}_1\bar \ell f_3} y^\sigma_{s_2 \bar f_3\ell} - y^{\bar\sigma}_{s_2\bar \ell f_3} y^\sigma_{\bar{s}_1 \bar f_3\ell} \right)
    F_S^B(x^{f_1}_{s_1},x^{s_1}_{s_2},x^{f_3}_{s_1})  
    \Bigg\}\\
    +& \sum\limits_{s_1s_2f_1Z}  \frac{g^\sigma_{Z\bar \ell\ell}}{M_Z^2}
    y^L_{s_2 \bar f_1d_j}y^R_{\bar{s}_1\bar d_i f_1}
    \left( \delta_{s_1s_2}g^L_{Z\bar d_jd_j} 
      +g_{Zs_1 \bar{s}_2}  \right )
    F^Z_S(x^{f_1}_{s_1},x^{f_1}_{s_2})  \\
+& \sum\limits_{f_1f_2s_1Z} \frac{g^\sigma_{Z\bar \ell\ell}}{M_Z^2}
y^L_{s_1\bar f_1d_j} y^R_{\bar s_1\bar d_if_2}
\left( g^L_{Z\bar f_2f_1}F^Z_{S^\prime}(x^{f_1}_{s_1},x^{f_2}_{s_1})+
g^R_{Z\bar f_2f_1}F^Z_{S^{\prime\prime}}(x^{f_1}_{s_1},x^{f_2}_{s_1})  \right) \,,
  \end{aligned}
\end{equation}
where we in both cases we have a single box function that covers both fermion flow directions, albeit with a sign difference.

\subsubsection*{Derivation of the pure vector part}

In the following we will show how the combination of the results of Ref.~\cite{Brod:2019bro} with our calculation of the photon penguin will lead to gauge independent results for the Wilson coefficients.
Denoting the contribution of the photon penguin that involves a photon coupling to the internal fermion and vector boson by $F_{\gamma}$ and $F_{\gamma'}$, respectively, we write\footnote{The additional function argument $\xi$ indicates that the loop function is gauge dependent. In the actual calculation, we kept the full dependence on the gauge parameters $\xi_{v}$ for each heavy vector boson, as defined in Eq.~\eqref{eq:gaugefix_massive_vector}.}
\label{sec:deriv-pure-vect}
\begin{equation}
\label{eq:v-before}
\begin{split}
  v_{L \sigma}^{i j \ell} = & \hspace{-12pt} \sum_{Zf_1f_2v_1v_2} \hspace{-12pt}
  \frac{g^L_{\bar{v}_2\bar d_i f_2} g^L_{v_1\bar f_1 d_j} g^\sigma_{Z \bar \ell\ell}}{M_Z^2}
  \Big\{
  \delta_{v_1 v_2} \left[
    g_{Z\bar{f}_2f_1}^L F_V^Z\left(x_{v_1}^{f_1},x_{v_1}^{f_2}\right) +
    g_{Z\bar{f}_2f_1}^R F_{V'}^Z\left(x_{v_1}^{f_1},x_{v_1}^{f_2}\right)
  \right] \\
  & \qquad
  + \delta_{f_1 f_2} g_{Zv_2 \bar v_1 }
  \left[
    F_{V''}^Z \left(x^{f_0}_{v_1},x^{f_1}_{v_1},x^{v_1}_{v_2} \right)
    + \frac{M_Z^2}{M_{v_1}^2} F_{V''}^{(2)}\left(x^{f_0}_{v_1},x^{f_1}_{v_1},x^{v_1}_{v_2}, \xi \right)
  \right]
  \Big\} \\
  + & \sum_{f_1v_1}
  g^L_{\bar{v}_1\bar d_i f_1} g^L_{v_1\bar f_1 d_j} g_{\gamma \bar \ell\ell} \frac{1}{M_{v_1}^2}
  \left[
    g_{\gamma\bar{f}_1f_1} F_\gamma\left(x^{f_0}_{v_1},x^{f_1}_{v_1}\right) +
    g_{\gamma v_1\bar{v}_1} F_{\gamma'}\left(x^{f_0}_{v_1},x^{f_1}_{v_1}, \xi \right)
  \right] \\
  + & \hspace{-6pt} \sum_{f_1f_3v_1v_2} \hspace{-6pt}
  g_{\bar{v}_2\bar d_if_1}^{L} g_{v_1\bar{f}_1d_j}^{L} \frac{1}{M_{v_1}^2}
  \Big[
    g_{\bar{v}_1\bar \ell f_3}^{\sigma} g_{v_2\bar{f_3}\ell}^{\sigma}
    F_B^{L\sigma}\big(x^{f_0}_{v_1},x^{f_1}_{v_1},x^{v_1}_{v_2},x^{f_3}_{v_1}, \xi\big)
    \\ & \hspace{96pt}
    - g_{v_2\bar{\ell}f_3}^{\sigma} g_{\bar{v}_1\bar f_3 \ell}^{\sigma}
    F_{B'}^{L\sigma}\big(x^{f_0}_{v_1},x^{f_1}_{v_1},x^{v_1}_{v_2},x^{f_3}_{v_1}, \xi\big)
  \Big]
\end{split}
\end{equation}
where all functions are independent of the masses $M_Z$ arising from one-particle reducible diagrams involving neutral massive vector-particle propagators.
The functions $F_{V^{(\prime,\prime\prime)}}^Z$ have already been combined with the terms that originate from the off-diagonal field renormalisation, as described in Ref.~\cite{Brod:2019bro}.
This combination is essential to arrive at a gauge-independent result.
In this context it is interesting to note that we can further use the sum rules to write $F_{V}^Z$ in a simpler and more symmetric form. 
The combination $F_{V''}^Z + (M_Z^2/M_{v_1}^2) F_{V''}^{(2)}$ agrees with the $F_{V''}$ of Ref.~\cite{Brod:2019bro} in the limit of 't~Hooft-Feynman gauge; here the gauge-parameter dependent part has been split off into the loop function $F_{V''}^{(2)}$.
The dependence on the mass of the lightest fermion $f_0$ originates from the application of the generalised GIM mechanism, Eq.~\eqref{eq:generalised-gim}, to our result.
It implies that the functions $F_{V''}^{(2)}$ approach zero in the limit $m_{f_1} \to m_{f_0}$.
The functions $F_{\gamma}$ and $F_{\gamma'}$ have been calculated here for the first time, while the box functions $F^{L\sigma}_{B^{(\prime)}}$ and $F^{L\sigma}_{B^{(\prime)}}$ are related to the expressions of Ref.~\cite{Brod:2019bro} in the limit $\xi_v = 1$ in the following manner:
\begin{align}
  F_{B}^{LL}(\cdot) &= - f_d(\cdot) - f_{\tilde{d}}(\cdot)\,, &
  F_{B'}^{LL}(\cdot) &= - f_d(\cdot) - 4 f_{\tilde{d}}(\cdot)\,,\\
  F_{B}^{LR}(\cdot) &= - f_d(\cdot) - 4 f_{\tilde{d}}(\cdot)\,, &
  F_{B'}^{LR}(\cdot) &= - f_d(\cdot) - f_{\tilde{d}}(\cdot) \,,
\end{align}
where
\begin{equation}
  \label{eq:6}
  \begin{split}
    f_d(x^{f_0}_{v_1},x^{f_1}_{v_1},x^{v_1}_{v_2},x^{f_3}_{v_1}, \xi) &= \frac{m_{f_1}^2m_{f_3}^2}{M_{v_2}^2}
                    \bigg\{
                    \frac{1}{4} \tilde{D}_0\big( m_{f_1},m_{f_3},m_{v_1},m_{v_2} \big) 
                    \\ &\quad -
                    (M_{v_1}^2+M_{v_2}^2) D_0\big( m_{f_1},m_{f_3},m_{v_1},m_{v_2}, \xi \big)
                    \bigg\} - (m_{f_1}\to m_{f_0})\,, \\
                    f_{\tilde{d}}(x^{f_0}_{v_1},x^{f_1}_{v_1},x^{v_1}_{v_2},x^{f_3}_{v_1}) =&
                    M_{v_1}^2 \tilde{D}_0\big( m_{f_1},m_{f_3},m_{v_1},m_{v_2} \big) - (m_{f_1}\to m_{f_0})\,.
  \end{split}
\end{equation}
For an arbitrary gauge-fixing parameters $\xi_v$, only the $f_d$ function contains $\xi_v$-dependent terms.
To combine the penguin and box contributions of (\ref{eq:v-before}) we specify the sum rule~\eqref{eq:unitarity_sumrule} to the interaction of leptons with vector bosons,
\begin{equation}
  \label{eq:4}
  \sum_{Z}\, g_{Z \bar \ell\ell}^{\sigma} g_{Z v_2 \bar{v}_1} =
  - \delta_{\bar{v}_1 v_2} g_{\gamma \bar \ell\ell}^{\sigma} g_{\gamma v_2 \bar{v}_1}
  - \sum_{f_3} \, \big( g_{\bar{v}_1 \bar \ell f_3}^{\sigma}
  g_{v_2 \bar f_3 \ell}^{\sigma} - g_{v_2 \bar \ell f_3}^{\sigma}
  g_{\bar{v}_1 \bar f_3 \ell}^{\sigma} \big) \,,
\end{equation}
which allows us to identify
\begin{equation}
  \label{eq:5}
  F_V^{\gamma Z} (x_{v_1}^{f_1}) =
  g_{\gamma\bar{f}_1f_1} F_\gamma\left(x_{v_1}^{f_1},x_{v_1}^{f_2}\right) +
  g_{\gamma v_1\bar{v}_1}
  \left[
    F_{\gamma'}\left(x^{f_0}_{v_1},x^{f_1}_{v_1}, \xi\right) -
    F_{V''}^{(2)}\left(x^{f_0}_{v_1},x^{f_1}_{v_1}, 1, \xi \right)
  \right]
\end{equation}
and
\begin{equation}
  F_V^{\sigma,B^{(\prime)} Z}(x^{f_0}_{v_1},x^{f_1}_{v_1},x^{v_1}_{v_2},x^{f_3}_{v_1}) =
  F_{B^{(\prime)}}^{L\sigma} (x^{f_0}_{v_1},x^{f_1}_{v_1},x^{v_1}_{v_2},x^{f_3}_{v_1}, \xi) -
    F_{V''}^{(2)} (x^{f_0}_{v_1},x^{f_1}_{v_1},x^{v_1}_{v_2},x^{f_3}_{v_1}, \xi ) \,.
\end{equation}
Using the explicit form of the loop functions, it can then be shown that the resulting expressions are independent of the gauge-fixing parameter.

\section{Applications to Beyond the Standard Model Phenomenology}
\label{sec:pheno-applications}

To exemplify our formalism we will apply it to models of new physics that address the current rare $B$-decay anomalies. 
In this context, it is standard to write vector and axial-vector current operators; the Wilson coefficients of this effective Lagrangian,
\begin{equation}
\label{eq:c9c10}
\mathcal{L}_{\mathrm{eff}} = \frac{1}{16\pi^2}
\left\{
  C_9^{\ell} (\bar{s} \gamma_{\mu} P_L b) (\bar{\ell} \gamma^{\mu} \ell) +
  C_{10}^{\ell} (\bar{s} \gamma_{\mu} P_L b) (\bar{\ell} \gamma^{\mu} \gamma_5 \ell)
\right\}\,,
\end{equation}
are related to our coefficients of Eq.~(\ref{eq:CVLL-formula}) via the linear transformation 
\begin{equation}
  \label{eq:basis-trafo}
  C_{9/10}^{\ell} = \frac{1}{2}
  \left( \tilde{C}_{L R}^{23 \ell} \pm \tilde{C}_{L L}^{23 \ell} \right) \,.
\end{equation}
The relation for the operators involving right-handed quarks can be inferred from the above relation, by replacing 
${C}_{L \sigma}^{23 \ell} \to {C}_{R\sigma}^{23 \ell}$\,.

If we are interested in deviations from the standard model background, we have to subtract the standard model one-loop contribution from our complete new-physics calculation; hence, we define
\begin{equation}
\label{eq:delta-c9-and-10}
C_{9/10}^{\ell\, \mathrm{NP}} = C_{9/10}^{\ell} - C_{9/10}^{\ell\, \mathrm{SM}}\,.
\end{equation}
The standard model contribution follows directly from the vector contribution of Eq.~(\ref{eq:cv-vector}) and reads
\begin{equation}
\label{eq:sm-c9-c10}
\begin{split}
C_{9}^{\ell\, \mathrm{SM}} &=  \frac{e^2 G_F V_{ts}^{*}V_{tb}}{\sqrt{2}}
\left\{
\frac{1}{s_W^2} F_V^{L,B Z}(0,x^t_W,1,0)
- 4F_V^{\gamma Z}(0,x^t_W)
\right\} \,, \\
C_{10}^{\ell\, \mathrm{SM}} &= - \frac{e^2 G_F V_{ts}^{*}V_{tb}}{\sqrt{2} s_W^2} F_V^{L,B Z}(0,x^t_W,1,0) \,, \\
\end{split}
\end{equation}
where we have used the fact that $F_{V^{(\prime)}}^Z (x,x) = 0 = F_{V''}^Z(x,y,1)$. 

\subsection{A $Z'$-Model with flavour off-diagonal couplings}
\label{sec:pheno:vector-fermion}

To demonstrate the utility of the expressions derived in
Sec.~\ref{sec:wilcos}, we begin by applying them to a simple
model~\cite{Kamenik:2017tnu} developed to address the $b\to s\ell\ell$
lepton flavour non-universality anomaly.
The model consists of a vector-like quark with up-type quantum numbers
which is additionally charged under a hidden $U(1)^\prime$ gauge group
spontaneously broken by the vacuum expectation value of a scalar
field~$\Phi$.
The relevant couplings of the mass eigenstates to the gauge bosons are given by
\begin{equation}
\begin{split}
\mathcal{L}_{\text {int }} \supset
&-\frac{e}{\sqrt{2}s_W} V_{t i}\left[\left(c_{L} \bar{t}+s_{L} \bar{T}\right)
  \slashed{W}^{+} P_{L} d_{i}\right]+\mathrm{h.c.} \\
&-\frac{e}{2 c_{W}s_W}\left[\left(c_{L} \bar{t}+s_{L} \bar{T}\right)
  \slashed{Z} P_{L}\left(c_{L} t+s_{L} T\right)
  - \frac{4}{3} s_{W}^{2} \big(\bar t \slashed{Z} t
+ \bar T \slashed{Z} T \big) \right]\\
&-\tilde{g} q^{\prime}\left[\left(s_{L} \bar{t}-c_{L} \bar{T}\right)
  \slashed{Z}^{\prime} P_{L}\left(s_{L} t-c_{L} T\right)
+ \left(s_{R} \bar{t}-c_{R} \bar{T}\right)
  \slashed{Z}^{\prime} P_{R}\left(s_{R} t-c_{R} T\right)\right] \\
  &- \tilde{g}\bar{\mu}\slashed{Z}^\prime\left(q'_{\mu,V}+q'_{\mu,A}\gamma_5\right)\mu
 \,.
\end{split}
\end{equation}
where, $s_{L/R}$ and $c_{L/R}$ are the sine and cosine of the left-/right-handed $t-T$ mixing angles and $\tilde{g}$ is the $U(1)'$ gauge coupling. The $U(1)'$ charge of the top partner, $T$, is denoted by $q'$ and that of the muon by $q_{\ell,V/A}'$ for the vectorial/axial couplings.
With these couplings, Eq.~\eqref{eq:cv-vector} directly gives the contribution to the Wilson coefficients which are
\begin{equation}
\begin{split}
  C_{9}^{\mu\, \mathrm{NP}} &=s_L^2 \left(
C_{9}^{\mu\,\text{SM}}(x^t_W \to x^T_W)-C_{9}^{\mu\,\text{SM}}\right)
-\frac{e^2 G_F V_{ts}^{*}V_{tb}}{\sqrt{2}}
s_L^2c_L^2\left\{\frac{1-4s_W^2}{s_W^2}  F_V^Z(x^t_W,x^T_W) \right.
\\
&+2\left.\tilde{g}^2 q' q'_{\mu,V} \frac{M_W^2}{M_{Z'}^2}
\Big(
2 F_V^Z(x^t_W,x^T_W) +
 \frac{s_R c_R}{s_L c_L} \left[ F_{V'}^Z(x^t_W,x^T_W) +  F_{V'}^Z(x^T_W,x^t_W) \right]
\Big)\right\}
\end{split}
\end{equation}
and
\begin{equation}
\begin{split}
C_{10}^{\mu\, \mathrm{NP}} &=s_L^2 \left(
C_{10}^{\mu\,\text{SM}}(x^t_W \to x^T_W)-C_{10}^{\mu\,\text{SM}}\right)
+\frac{e^2 G_F V_{ts}^{*}V_{tb}}{\sqrt{2}}
s_L^2c_L^2\left\{\frac{1}{s_W^2}  F_V^Z(x^t_W,x^T_W) \right.
\\
&-2\left.\tilde{g}^2 q' q'_{\mu,A} \frac{M_W^2}{M_{Z'}^2}
\Big(
2 F_V^Z(x^t_W,x^T_W) +
\frac{s_R c_R}{s_L c_L} \left[ F_{V'}^Z(x^t_W,x^T_W) +  F_{V'}^Z(x^T_W,x^t_W) \right]
\Big)\right\}\,,
\end{split}
\end{equation}
where we have subtracted the SM contribution.
To evade collider constraints, one furthermore assumes that $m_T \gg m_t$.
In this limit we find:
\begin{equation}
\label{eq:c910zprimenp}
C_{9/10}^{\mu, \mathrm{NP}} = \frac{s_R^2 }{2} q' q'_{\mu,V/A} \frac{m_t^2}{M_{Z'}^2} \frac{\tilde{g}^2}{e^2} 
\left\{ \frac{1}{2}\log \left(x_W^T\right) + \frac{1}{c_R^2} +\frac{3}{2(x_t-1)}-1-\frac{1}{2}\left(\frac{3 }{(x_t-1)^2} +1\right)\log (x_W^t)\right\}
\end{equation}
where the $\log (x_W^T)$ agrees with the result in Ref.~\cite{Kamenik:2017tnu}, while the remaining terms are new and reduce the contribution to both $C_9$ and $C_{10}$ by 13(7)\% for $m_T=1(10)$ TeV.

\subsection{A $U(1)_{L_\mu-L_\tau}$ model with Majorana fermions}
\label{sec:pheno:vector-fermion-scalar}

The gauged $U(1)_{L_\mu-L_\tau}$ model was originally proposed in Refs.~\cite{He:1990pn,He:1991qd} and has been studied extensively in the context of lepton universality violation.
Here we focus on the model of Ref.~\cite{Baek:2017sew} where an additional Dirac fermion $N$ and a coloured $SU(2)_L$-doublet scalar $\tilde q\equiv (\tilde u, \tilde d)^T$ with hypercharge $Y = 1/6$ are introduced that are all charged under the $L_\mu-L_\tau$ gauge group.
After spontaneous symmetry breaking the relevant interactions in terms of the mass eigenstates read
\begin{equation}\label{broken-lag}
\begin{split}
  \mathcal L_\text{int} \supset &
  -\frac{g_X Q}{2} \left(\overline N_-+\overline N_+\right) \slashed Z^\prime \left(N_-+N_+\right)  
  - \frac{1}{\sqrt 2}
  \left[\left( y^b_L\bar b_L + y^s_L\bar s_L \right)
    \tilde d\left(N_-+N_+\right)+\text{h.c.}\right]
  \\
  &- i
  \left(
    g_XQ Z^\prime_\mu +
    g \frac{3 - 2 s_W^2}{6 c_W} Z_\mu
  \right)
  \left[
    \tilde d \partial_\mu \tilde d^{\,c}
    - \left( \partial^\mu\tilde d \right) \tilde d^{\,c} 
  \right] - g_X \bar{\mu} \slashed{Z} \mu \,,
\end{split}
\end{equation}
where $N_{\pm}=\left(N\pm N^c\right)/\sqrt{2}$ is written in term of $N$ and its charge conjugated field $N^c$, $g_X$ is the $U(1)_{L_\mu-L_\tau}$ gauge coupling, $Q$ is the charge of $N$, and $y^{s/b}_L$ are the Yukawa couplings of the SM bottom and strange quarks to $\tilde{d}$.

The $Z'$ penguin does not involve any SM particles and is lepton universality violating by construction.
The complete one-loop new physics contributions to $C_9^\mu$ can be read off from Eq.~\eqref{eq:cv-scalar}.
Noting that the charge conjugated scalar $\tilde{d}^{\,c}$ contributes in the sum of (\ref{eq:cv-scalar}), we find
\begin{equation}
  \begin{split}
    C_{9}^{\mu\, \mathrm{NP}} &=
    \frac{g_X^2 Q y^b_Ly^s_L}{4 M_{Z'}^2}
    \left[
      \sum\limits_{f_1,f_2=N_\pm}
      \left\{
        F^{Z}_{S'}(x^{f_1}_{\tilde d},x^{f_2}_{\tilde d}) +
        F^{Z}_{S''}(x^{f_1}_{\tilde d},x^{f_2}_{\tilde d})       
      \right\}
      - 2 \sum\limits_{f_1=N_\pm} F^{Z}_S(x^{f_1}_{\tilde d})
    \right] \\
    & \quad -
    \frac{e^2 y^b_Ly^s_L}{m_{\tilde{d}}^2} \sum\limits_{f_1=N_\pm}
      F^{\gamma}_{S}(x^{f_1}_{\tilde d}) \,,
  \end{split}
\end{equation}
where the first line represents the $Z'$-penguin contribution and agrees with the results of Ref.~\cite{Baek:2017sew}.
The terms in the second line represent the lepton flavour universal new physics contribution to $C_9$ from the photon-penguin and is new.
Note that the photon-penguin decouples faster than the $Z'$~penguin in the limit of large scalar mass $m_{\tilde{d}}$.
The $Z$ coupling to the down quarks cancels with the $Z$ couplings to $\tilde{d}^{\,c}$ in (\ref{eq:cv-scalar}) so that the $Z$-penguin contribution cancels.
The contribution to $C_{7,bs}^{\text{NP}}$ can be calculated from the general formula 
(\ref{eq:dipole:formula}) and is given by
\begin{equation}
C_{7,bs}^{\text{NP}} = \frac{1}{m_b}
\frac{y_L^by_L^s}{2 m_{\tilde d}^2}
\sum_{f_1=N_{\pm}}
F^{Z}_{S'}(x^{f_1}_{\tilde d})\,,
\end{equation}
where the operator $O_{7}^{bs}$ is defined in footnote~\ref{footnote:operators}. Note that only one of the terms is present since $N$ is electrically neutral and therefore only the charged scalar, $\tilde{d}^{\,c}$, contributes.

\subsection{A model with vector-like fermions and neutral scalars}
\label{sec:pheno:fermion-scalar}

To give another application of our results, we consider a model that consists of $SU(2)_L$ doublet vector-like quarks and leptons in addition to one or two complex scalars that are neutral under the SM gauge group~\cite{Grinstein:2018fgb}.
The interaction Lagrangian of interest reads
\begin{equation}
\begin{split}
\mathcal{L} \supset \frac{1}{\sqrt{2}}&\left\{
\left[y_{\Phi_L\bar{b}\Psi_Q}^{R}\,\Phi_L+y_{\Phi_H\bar{b}\Psi_Q}^{R}\,\Phi_H\right]\bar{b}P_R\Psi_Q +
\left[y_{\Phi_L\bar{s}\Psi_Q}^{R}\,\Phi_L+y_{\Phi_H\bar{s}\Psi_Q}^{R}\,\Phi_H\right]\bar{s}P_R\Psi_Q\right.\\
&\quad+\left. \left[y_{\Phi_L\bar{\ell}\Psi_\ell}^{R}\,\Phi_L+y_{\Phi_H\bar{\ell}\Psi_Q}^{R}\,\Phi_H\right]\bar{\ell}P_R\Psi_\ell + \mbox{h.c.}\,.
\right\}
\end{split}
\end{equation}
Hermitian conjugation gives the left-handed Yukawa couplings, $y^L$, which are related to the right-handed ones via
\begin{equation}
y_{\bar{\Phi}\bar{\Psi}f}^L = \left(y_{\Phi\bar{f}\Psi}^R\right)^*,
\end{equation}
where $\Phi\in\left\{\Phi_L,\Phi_H\right\}$, $\Psi\in\left\{\Psi_Q,\Psi_\ell\right\}$, and $f\in\left\{b,s,\ell\right\}$ as applicable. The expressions for the Yukawa couplings can be read off from Ref.~\cite{Grinstein:2018fgb} and we omit writing them explicitly.
The NP contribution to $C_9$ and $C_{10}$ are, then,
\begin{equation}
C_{9/10}^{\mu,\text{NP}} = 
\frac{1}{2}\left( s_{L R}^{23\mu} \pm s_{L L}^{23\mu} \right)\,,
\end{equation}
and, from Eq.~\eqref{eq:cv-scalar}, we have
\begin{equation}
\begin{split}
\left.s_{LR}^{23\mu}\right|_\text{box} &= 0\\
\left.s_{L L}^{23\mu}\right|_\text{box}
&= \frac{1}{4M_{\Phi_L}^2}\,y^L_{\Phi_L^* \bar\Psi_Q b}\,y^R_{\Phi_L\bar s \Psi_Q}
\left|y^{R}_{\Phi_L\bar \mu \Psi_\ell}\right|^2
F_S^B(x^{\Psi_Q}_{\Phi_H},1,x^{\Psi_\ell}_{\Phi_L})\\
&+  \frac{1}{4M_{\Phi_L}^2}\,y^L_{\Phi_H^* \bar\Psi_Q b} \,y^R_{\Phi_H\bar s \Psi_Q}
\left|y^{R}_{\Phi_H\bar \mu \Psi_\ell}\right|^2
F_S^B(x^{\Psi_Q}_{\Phi_H},1,x^{\Psi_\ell}_{\Phi_H})\\
&+  \frac{1}{4M_{\Phi_L}^2}\,y^L_{\Phi_L^* \bar\Psi_Q b} \,y^R_{\Phi_H\bar s \Psi_Q}
\left(
y^{R}_{\Phi_L\bar \mu \Psi_\ell} \, y^L_{\Phi_H^* \bar\Psi_\ell\mu} 
\right)
F_S^B(x^{\Psi_Q}_{\Phi_L},x^{\Phi_L}_{\Phi_H},x^{\Psi_\ell}_{\Phi_L})\\
&+  \frac{1}{4M_{\Phi_H}^2}\,y^L_{\Phi_H^* \bar\Psi_Q b} \,y^R_{\Phi_L\bar s \Psi_Q}
\left( y^{R}_{\Phi_H\bar \mu \Psi_\ell} \, y^L_{\Phi_L^* \bar\Psi_\ell\mu} 
\right)
F_S^B(x^{\Psi_Q}_{\Phi_H},x^{\Phi_H}_{\Phi_L},x^{\Psi_\ell}_{\Phi_H})
\,.
\end{split}
\end{equation}
Note that $C_9^\ell$ receives a lepton-flavour-universal contribution from the photon penguin. This contribution breaks the relation $C_9 = -C_{10}$ but it is suppressed by fermion masses and is therefore subleading in the limit where the scalars are lighter. 
Substituting the couplings from Ref.~\cite{Grinstein:2018fgb} and translating the box functions, $F_S^B$, into their $G$ functions gives perfect agreement with their result.

\section{Summary and conclusions}
\label{sec:summary-conclusions}

In this work, we have presented finite and manifestly gauge-invariant matching contributions at the one-loop level onto the weak effective Lagrangian in generic extensions of the SM. That is, we add to its field content any number of massive vector bosons, physical scalars, and fermions. 
For a given field content, only a minimal number of couplings needs to be specified because perturbative unitarity of the S-matrix implies that not all couplings can be independent.
The constraints on the couplings are codified in the sum rules that arise from Slavnov-Taylor identities which are in turn obtained from the invariance of appropriate Green's functions under BRST transformations.

The main results of this paper, the sum rules on the additional couplings and the finite and gauge-invariant one-loop contribution, are implemented in a \texttt{Mathematica} package available for download from
\begin{center}
	\url{https://wellput.github.io}\,.
\end{center}
This package contains an example file that includes the SM contribution to the operators considered in this paper along with all three extensions discussed in Sec.~\ref{sec:pheno-applications}.
Specifically, we considered three classes of extensions that demonstrate the three types of contributions in Eqs.~\eqref{eq:cv-vector}, \eqref{eq:cm-mixed}, and \eqref{eq:cv-scalar} corresponding to the addition of massive vectors and fermion (Sec.~\ref{sec:pheno:vector-fermion}), vectors, fermions, and scalars (Sec.~\ref{sec:pheno:vector-fermion-scalar}), and scalars and fermions (Sec.~\ref{sec:pheno:fermion-scalar}), respectively.

Finally, the scope of this paper was to implement the matching onto the $|\Delta F|=1$ dipole and current-current weak effective Lagrangian Wilson coefficients. 
The extension to flavour-conserving magnetic and electric dipole operators and to dimension-six scalar operators is already work-in-progress and will appear in the near future.

\section*{Acknowledgments}
MG is supported by the UK STFC under Consolidated Grant
ST/T000988/1 and also acknowledges support from COST Action CA16201
PARTICLEFACE.
JB acknowledges support by DOE grant DE-SC0011784.
This work was also supported by the Deutsche Forschungsgemeinschaft (DFG, German Research Foundation) under Germany's Excellence Strategy - EXC 2121 ``Quantum Universe'' - 390833306 and the Aspen Center for Physics, which is supported by National Science Foundation grant PHY-1607611.
UM is supported by the Bolashak International Scholarship Programme.

\appendix
\section{Loop Functions}
\label{sec:loop-functions}

In this appendix we collect the analytical expressions of all loop functions that appear in the final results for the renormalised Wilson coefficients.
These functions depend on the masses of the particles inside the respective loop diagrams and on their electromagnetic charges.

\subsection{Loop Functions for the Dipole Coefficients}
\label{sec:loop-functions-dipole}

In the limit where no particles are much lighter than the matching scale, we find  the functions involving scalars,
\begin{equation}
\begin{split}
F^d_{S}(x) &=
Q_{s_1}\left(\frac{x+1}{4 (x-1)^2}-\frac{x \log (x)}{2(x-1)^3}\right) +
Q_{f_1} \left(\frac{\log (x)}{2(x-1)^3}+\frac{x-3}{4 (x-1)^2}\right) \,,\\
F^d_{S^\prime}(x) &=
Q_{s_1}\left(\frac{2 x^2+5 x-1}{24 (x-1)^3}-\frac{x^2 \log (x)}{4(x-1)^4}\right) +
Q_{f_1} \left(\frac{x \log (x)}{4(x-1)^4}+\frac{x^2-5 x-2}{24 (x-1)^3}\right) \,,
\end{split}
\end{equation}
and vectors,
\begin{equation}
\begin{split}
F^d_V(x_0,x) &= f^d_V(x) - f^d_V(x_0) \,,\\
f^d_V(x) &=
Q_{v_1} \left(\frac{11 x^2-7 x+2}{8(x-1)^3}-\frac{3 x^3 \log(x)}{4 (x-1)^4}\right)+
Q_{f_1} \left(\frac{3 x^2 \log (x)}{4(x-1)^4}-\frac{2x^2+5 x-1}{8(x-1)^3}\right) \,, \\
F^d_{V^\prime}(x) &=
Q_{v_1} \left(\frac{3 x^2 \log (x)}{2(x-1)^3}+\frac{x^2-11x+4}{4 (x-1)^2}\right)+
Q_{f_1} \left(\frac{x^2+x+4}{4 (x-1)^2}-\frac{3 x \log(x)}{2 (x-1)^3}\right)\,,
\end{split}
\end{equation}
that contribute to the Wilson coefficient of the dipole operator in Eq.~(\ref{eq:dipole:formula}).
As stated above, our results agree with Ref.~\cite{Lavoura:2003xp} after employing the relevant unitarity sum rule.

\subsubsection*{Limit of light internal particles}

Light internal particles can in principle give a contribution from the effective theory side of the matching equation. 
The scalar loop functions that multiplies Yukawa couplings of the same chirality must contain an odd number of chirality flips as explained above.
This implies that the infrared logarithm $\sqrt{x} \log(x)$ vanishes in the limit $x \to 0$.
Since we work at dimension five for our dipole operators, the effective theory contribution is vanishing in this limit and we do not have to consider the scalar functions further. 
The vector contributions of the dipole operator have no infrared logarithm in the limit of the lightest internal fermion mass tending to zero.
Since $F_{V'}^d$ is multiplied with the internal fermion mass we only need to consider the limit $x_0 \to 0$ for $F_{V}^d$ and find:
\begin{equation}
F^d_V(0,x) =
x \Bigg\{
Q_{v_1} \left(-\frac{3 x^2 \log(x)}{4 (x-1)^4}+\frac{2 x^2+5 x-1}{8(x-1)^3}\right)+
  Q_{f_1} \left(\frac{3 x \log (x)}{4(x-1)^4}+\frac{x^2-5 x-2}{8(x-1)^3}\right) \Bigg \}\,.
\end{equation}

\subsection{Loop Functions for the Neutral-Current Operators}
\label{sec:loop-functions-vector}

We first give the functions that contribute to the Wilson coefficient of the neutral-current operators in the scenario where no light internal particles are in the loop.
We start with the first term in Eq.~(\ref{eq:CVLL-formula}) that comprises the contributions of internal vector bosons and fermions.
We find the following gauge-invariant combination of the photon penguin and the $Z$ Penguin
\begin{equation}\label{fvaz-massive}
  F_V^{\gamma Z}(x_0,x) = f_V^{\gamma Z}(x) - f_V^{\gamma Z}(x_0) \,,
\end{equation}
where
\begin{equation}
\begin{split}
 f_V^{\gamma Z}(x)
 &= Q_{f_1} \left(\frac{14 x^2-21 x+1}{12 (x-1)^3}
    +\frac{\left(-9 x^2+16 x-4\right)}{6 (x-1)^4}  \log(x)\right)\\
 & \quad + Q_{v_1} \left( \frac{6 x^4-18 x^3-32 x^2+87 x-37}{12 (x-1)^3} + 
   \frac{x \left(8 x^3-2 x^2-15 x+6\right)}{6 (x-1)^4}\log(x)\right)\,.
\end{split}
\end{equation}
The terms proportional to $Q_f$ and $Q_v$ originate from the photon penguin and the combination of the $Z$-penguin and the photon penguin, respectively.
The remaining loop functions involving vector bosons are
\begin{equation}
  F_{V''}^Z(x_0,x,y) \equiv f_{V''}^Z(x,y) - f_{V''}^Z(x_0,y) \,,
\end{equation}
with
\begin{equation}
\begin{split}
  f_{V''}^Z(x,y)
&= -\frac{x (y-1) \left(3 x^2 (y-1) y-10 x y+4\right) }{4 (x-1) (x y-1)^2}\log (x)+
\frac{x \left(2 x y^2-2x y+y+5\right)}{4 x y-4} \\[0.5em]
& \quad +\frac{x y \left(x \left(-4 y^2-5 y+3\right)+y+5\right) }{4 (y-1) (xy-1)^2}\log (y)\,,
\end{split}
\end{equation}
as well as
\begin{equation}
  F_V^{L,BZ}(x_0, x, y, z) \equiv f_V^{L,BZ}(x, y, z) - f_V^{L,BZ}(x_0, y, z)\,,
\end{equation}
with
\begin{equation}
\begin{split}
  f_V^{L,BZ}(x, y, z)
& =\frac{x y \left(3 x^2 y (x y+x-2)-(x-1) z (x y (x y-2)+4)\right)}{4 (x-1) (x y-1)^2
   (x-z)}\log (x)\\[0.5em]
& \quad +\frac{3 x y^2  (x+(y-1) z-1)}{4 (y-1) (x y-1)^2 (y z-1)}\log (y)\\[0.5em]
& \quad +\frac{x y z (z (y(z-4)-4)+4)}{4 (z-1) (x-z) (y z-1)}\log (z)+\frac{x y (2 x y-5)}{4 x y-4}\,,
\end{split}
\end{equation}
and 
\begin{equation}
  F_V^{R,BZ}(x_0, x, y, z) \equiv f_V^{R,BZ}(x, y, z) - f_V^{R,BZ}(x_0, y, z)\,,
\end{equation}
with
\begin{equation}
\begin{split}
  f_V^{R,BZ}(x, y, z)
& =\frac{x y  (3x (x y (x y+x-6)+4)-(x-1) z (x y (x y-2)+4))}{4 (x-1) (x y-1)^2 (x-z)}\log (x) \\[0.5em]
& \quad +\frac{3 x y^2 (-4 x y+x+(y-1) z+3)}{4 (y-1) (x y-1)^2 (y z-1)}\log (y)\\[0.5em]
& \quad +\frac{x y (z-4) z(y z-4) \log (z)}{4 (z-1) (x-z) (y z-1)}+\frac{x y (2 x y-5)}{4 x y-4}\,.
\end{split}
\end{equation}
In addition, we have
\begin{equation}
F_{V}^{Z} (x, y)
 =   \frac{x y }{x-y}\log\left(\frac{x}{y}\right)
   - \frac{x+y}{2}
\end{equation}
and
\begin{equation}
\begin{split}
F_{V'}^{Z} (x, y) & =
 \sqrt{xy} \bigg[
 \frac{y-4}{2 \left(y-1\right)}-\frac{\left(x-4\right) x}{2\left(x-1\right) \left(x-y\right)}
 \log \left(x\right) \\[0.5em]
& \qquad \qquad
  +\frac{\left(x \left(\left(y-2\right)y+4\right)-3 y^2\right)}{2 \left(x-y\right)\left(y-1\right)^2}
  \log \left(y\right) \bigg]\,.
\end{split}
\end{equation}
Moerover, we have the relations
\begin{align}
F_V^{L,B^\prime Z}(x, y, z) = F_V^{R,BZ}(x, y, z), && F_V^{R,B^\prime Z}(x, y, z) = F_V^{L,BZ}(x, y, z).
\end{align}
The loop functions involving scalar particles are
\begin{align}
\begin{split}
F_S^\gamma(x) & =
    Q_s \left(\frac{11 x^2-7 x+2}{36
   (x-1)^3}-\frac{x^3 \log (x)}{6
   (x-1)^4}\right) \\[0.5em]
   & \quad + Q_{f_1} \left(\frac{7x^2-29 x+16}{36 (x-1)^3}+\frac{(3 x-2) \log (x)}{6 (x-1)^4}\right)\,,
\end{split}\\[0.5em]
F^Z_S(x,y) & =
   \frac{1-2 y}{2 \left(y-1\right)}-\frac{y \log
   \left(x\right)}{2 \left(x-1\right)
   \left(x-y\right)}+\frac{\left(x-1\right) y \log
   \left(y\right)}{2 \left(x-y\right)
   \left(y-1\right)^2}\,,\\[0.5em]
F^Z_{S^\prime}(x,y) & =
   \frac{\sqrt{xy}}{(x-y)}
   \left( \frac{x\log(x)}{x-1}-\frac{y\log(y)}{y-1} \right)\,,\\[0.5em]
F^Z_{S^{\prime\prime}}(x,y) & = 
  \frac{y}{2 \left(y-1\right)}-\frac{x^2 \log
   \left(x\right)}{2 \left(x-1\right)
   \left(x-y\right)}+\frac{y \left(x
   \left(y-2\right)+y\right) \log \left(y\right)}{2
   \left(x-y\right) \left(y-1\right)^2}\,,\\[0.5em]
\begin{split}
F^B_S\left(x,y,z\right) & = 
      \frac{x^2 y \log (x)}{4 (x-1) (x y-1) (x-z)}\\[0.5em]
  & \quad + \frac{y z^2 \log (z)}{4 (z-1) (z-x) (y
      z-1)}+\frac{y \log (y)}{4 (y-1) (x y-1) (y z-1)}\,,
\end{split}
\end{align}
while the loop functions with both vectors and scalars are
\begin{align}
\begin{split}
F^B_{VS}(x,y,z) & =
  \sqrt{xz}\Bigg[
  \frac{x (x y-4) \log (x)}{4 (x-1) (x y-1) (x-z)}\\[0.5em]
& \qquad \qquad -\frac{3 y \log (y)}{4 (y-1) (x y-1) (y
   z-1)}-\frac{z (y z-4) \log (z)}{4 (z-1) (x-z) (y z-1)}\Bigg]\,,
\end{split}\\[0.5em]
F^Z_{VS}(x,y) & =
  \sqrt{y}\left[
  -\frac{\left(y-4 x\right) \log \left(x\right)}{4 \left(x-1\right)
  \left(x-y\right)}+\frac{y \left(x+2 y-3\right) \log \left(y\right)}{4
  \left(y-1\right){}^2 \left(y-x\right)}+\frac{5-4 y}{4 \left(y-1\right)} \right]\,,\\[0.5em]
F^Z_{VS^\prime}(x,y) & =
   \sqrt{y}\left[
   \frac{x \left(4 x-y-3\right) \log \left(x\right)}{4 \left(x-1\right){}^2
   \left(x-y\right)}-\frac{3 x \log \left(y\right)}{4 \left(y-1\right)
   \left(x-y\right)}+\frac{1-2 x}{4 \left(x-1\right)}\right]\,.
\end{align}

\subsubsection*{Internal light fermion}

In the limit of a light internal fermion the function $F_V^{\gamma Z}(x_0, x)$ in Eq.~(\ref{fvaz-massive}) exhibits an infrared logarithm $\log x_0$.
This logarithm is cancelled through the effective theory contribution of the light fermion.
A tree-level matching of the vector boson contributions will generate a four-fermion Wilson-coefficient that has a non-vanishing one-loop matrix element whose projection $\delta r_{\sigma\sigma'}^{ij \ell}$ onto the tree-level matrix element of the neutral current operator of Eq.~(\ref{eq:lag:5flavour}) reads
\begin{equation}
\label{eq:delta-r}
\delta r_{\sigma\sigma'}^{ij \ell} = \sum\limits_{v_1 f_1}
\frac{g^\sigma_{\bar{v}_1\bar d_i f_1} g^\sigma_{v_1\bar f_1 d_j}}{M_{v_1}^2}
\frac{e^2 Q_{\ell} Q_f}{16 \pi^2} 
\left( \frac{2}{3} - \frac{2}{3} \log \frac{\mu^2}{m^2} \right)\,,
\end{equation}
if we keep the dependence on the light fermion mass to regularise the infrared divergence and include the operator mixing of the tree-level operator to renormalise the ultraviolet pole.
Subtracting this effective theory contribution from our full theory result, the light mass dependence will cancel out and obtain the matching corrections in the limit of light internal fermion masses.
\begin{equation}
\begin{split}
F_V^{\gamma Z}(0,x) &= 
Q_{f_1} \left(\frac{2}{3}\log\left(\frac{\mu^2}{M_{v_1}^2}\right) +
\frac{x\left(x^2+11 x-18\right)}{(x-1)^3}+\frac{\left(-9 x^2+16 x-4\right)}{6 (x-1)^4}\log(x)\right)\\
& \quad +Q_{v_1} \left(\frac{x \left(6 x^3-41 x^2+77 x-48\right)}{12 (x-1)^3}+\frac{x \left(10 x^3-22 x^2+9x+6\right) \log (x)}{6 (x-1)^4}\right)\,.
\end{split}
\end{equation}

\section{Additional Sum Rules}
\label{sec:relevant-sum-rules}

The following two additional sum rules are required to obtain a finite and gauge-independent result for the $Z$ penguin:
\begin{align}
\begin{split}\label{eq:gauge_mass_sumrule}
  \isum{s_1} \, g_{v_1v_2 \bar s_1 } y_{s_1 \bar f_1 f_2}^{\sigma} & =
  \isum{v_3} \, \tfrac{M_{v_1}^2-M_{v_2}^2}{M_{v_3}^2} \,
  g_{v_1v_2 \bar v_3 } \Big( m_{f_1} g_{v_3 \bar f_1 f_2}^{\sigma} -
  g_{v_3 \bar f_1 f_2}^{\bar\sigma} m_{f_2} \Big)\\
  & \quad + \isum{f_3} \Big( - m_{f_1} \big(
  g_{v_2 \bar f_1 f_3}^{\sigma} g_{v_1 \bar f_3 f_2}^{\sigma} +
  g_{v_1 \bar f_1 f_3}^{\sigma} g_{v_2 \bar f_3 f_2}^{\sigma} \big)\\
  & \qquad \qquad - m_{f_2} \big( g_{v_2 \bar f_1 f_3}^{\bar\sigma}
  g_{v_1 \bar f_3 f_2}^{\bar\sigma} + g_{v_1 \bar f_1 f_3}^{\bar\sigma}
  g_{v_2 \bar f_3 f_2}^{\bar\sigma} \big) \\
  & \qquad \qquad + 2\, m_{f_3} \big( g_{v_2 \bar f_1 f_3}^{\bar\sigma}
  g_{v_1 \bar f_3 f_2}^{\sigma} + g_{v_1 \bar f_1 f_3}^{\bar\sigma}
  g_{v_2 \bar f_3 f_2}^{\sigma} \big) \Big) \,,
\end{split}\\[1em]
\begin{split}\label{eq:yukawa_sumrule}
  \isum{s_1} \, g_{v_1s_2 \bar s_1 } y_{s_1 \bar f_1 f_2}^{\sigma} & =
  - \isum{v_3} \, \tfrac{1}{2 M_{v_3}^2} g_{v_1 \bar v_3 s_2} \Big(
  m_{f_1} g_{v_3 \bar f_1 f_2}^{\sigma} - g_{v_3 \bar f_1 f_2}^{\bar\sigma}
  m_{f_2} \Big)\\
  & \quad + \isum{f_3} \Big( g_{v_1 \bar f_1 f_3}^{\bar\sigma}
  y_{s_2 \bar f_3 f_2}^{\sigma} - y_{s_2 \bar f_1 f_3}^{\sigma}
  g_{v_1 \bar f_3 f_2}^{\sigma} \Big) \,.
\end{split}
\end{align}

\addcontentsline{toc}{section}{References}
\bibliographystyle{JHEP}
\bibliography{references}

\end{document}